\documentclass[superscriptaddress,rmp]{revtex4}
\usepackage{amsmath}
\usepackage{amsfonts}
\usepackage{amssymb}
\usepackage{epsfig}

\newcommand{\ra}{\rangle}
\newcommand{\la}{\langle}

\begin{document}

\title{Current Fluctuations and Statistics During a Large Deviation Event in an Exactly-Solvable Transport Model}

\author{Pablo I. Hurtado}
\email{phurtado@onsager.ugr.es}
\author{Pedro L. Garrido}
\email{garrido@onsager.ugr.es}
\affiliation{Departamento de Electromagnetismo y F\'{\i}sica de la Materia, 
and Instituto Carlos I \\
de F\'{\i}sica Te\'orica y Computacional, Universidad de Granada, 
Granada 18071, Spain}

\date{\today}

\begin{abstract}
We study the distribution of the time-integrated current in an exactly-solvable toy model of 
heat conduction, both analytically and numerically. The simplicity of the model allows us to derive the full
current large deviation function and the system statistics during a large deviation event. In this way we unveil a relation
between system statistics at the end of a large deviation event and for intermediate times. Midtime statistics 
is independent of the sign of the current, a reflection of the time-reversal symmetry of microscopic dynamics, while endtime 
statistics do depend on the current sign, and also on its microscopic definition. 
We compare our exact results with simulations based on the direct evaluation of large deviation functions, 
analyzing the finite-size corrections of this simulation method and deriving detailed bounds for its applicability.
We also show how the Gallavotti-Cohen fluctuation theorem can be used to determine the range of validity of simulation results.
\end{abstract}


\maketitle

\section{Introduction}

Many nonequilibrium systems typically exhibit currents of different observables (e.g., mass, energy, spin, etc.)
which characterize their macroscopic behavior. These currents fluctuate, and the (large deviation) functional
which determines the probability of these fluctuations is a natural candidate to generalize the concept 
of \emph{free energy} to nonequilibrium systems. In this way, understanding how microscopic dynamics
determine the long-time averages of these currents and their fluctuations is one of the main objectives
of nonequilibrium statistical physics \cite{BD,Bertini,B2,Livi,Dhar,we,Derrida,GC,LS,K}.
An important step in this direction has been the development of fluctuation theorems \cite{GC,LS,K}, which relate the
probability of forward and backward currents reflecting the time-reversal symmetry of microscopic dynamics. These are relations of the form
\begin{equation}
\frac{\text{P}(q,t)}{\text{P}(-q,t)}\sim \text{e}^{+t E q} \, ,
\label{fth}
\end{equation}
where $\text{P}(q,t)$ is the probability of observing a time-integrated current $Q_t=qt$ after a long time $t$, and $E$ is some constant such that 
$E q$ corresponds to the rate of entropy production \cite{Derrida,GC,LS,K,Rakos}. Despite this progress,
we still lack a general approach based on few simple principles to understand current statistics in nonequilibrium systems. 
This has triggered an intense research effort in recent years. For instance,
Bertini, De Sole, Gabrielli, Jona-Lasinio and Landim \cite{Bertini,B2} have recently introduced a hydrodynamic fluctuation theory (HFT)
to study both current and profile large dynamic fluctuations in nonequilibrium steady states, 
providing a variational principle which describes the most probable (possibly time-dependent) profile responsible of a given current fluctuation. 
Simultaneously, Bodineau and Derrida \cite{BD} have conjectured an additivity principle for current fluctuations which enables one to 
explicitely calculate the current distribution for 1D diffusive systems in contact with two boundary baths. 
This principle, which is equivalent within HFT to the hypothesis that the optimal profile is time-independent \cite{B2}, has been recently confirmed in a model of 
heat conduction \cite{HG}.

The predictions derived from the additivity principle and the hydrodynamic fluctuation theory 
have been tested in few models for which the exact current distribution can be calculated. For more general cases in which no exact solutions 
are available, new algorithms that allow the direct evaluation of large deviation functions in computer simulations have been recently proposed \cite{sim,sim2}. 
These methods are based on a modification of the microscopic dynamics, so that the rare events responsible of the current large deviation
are no longer rare. Though successful, the new algorithms seem to suffer from finite-size effects when measuring the probability of extreme current deviations \cite{HG}. 
Here again the existence of simplified models for which exact solutions can be obtained is essential to elucidate the origin and importance of these 
finite-size corrections in simulations. 

The aim of this paper is to present one of these simplified models, i.e. a toy model of heat transport between two thermal baths at different temperatures. 
The simplicity of this model allows us to gain a better understanding of:
(i) system statistics during a large deviation event and how it reflects the
time reversibility of microscopic dynamics, and (ii) finite-size corrections to the direct evaluation of large deviation functions.
The model and the calculation of its current large deviation function are
presented is Section II. The probabilities of having certain configuration 
at the end of a large deviation event and for intermediate times are calculated in Section III, while Section IV is
devoted to analyze their relation in terms of the reversibility of microscopic dynamics. Simulation results are described in Section V,
together with an analysis of finite-size corrections affecting the direct
evaluation of large deviation functions. This analysis provides a deep
understanding of the range of validity of this simulation method, which otherwise can be determined using the Gallavotti-Cohen fluctuation theorem. Finally, Section VI contains our conclusions.

\section{Model and Exact Solution}

Our toy model consists in a single lattice site characterized by an energy $e\in \mathbb{R}_+$. This site is coupled to two thermal baths 
at different temperatures, $T_L$ (left) and $T_R$ (right), and we assume $T_L>T_R$ without loss of generality. Dynamics is stochastic and proceeds as follows: 
(i) With equal probability, we randomly choose the system to interact with one of the heat baths, $T_L$ or $T_R$; (ii) a new energy $e'$ is drawn from the distribution 
$\beta_{\nu}\text{exp}(-\beta_{\nu}e')$ corresponding to the (inverse) temperature of the selected heat bath, $\beta_{\nu}=T_{\nu}^{-1}$, with $\nu=L,R$.  
The presence of a temperature gradient then forces the system out of equilibrium. The transition rate $U_{e' e}^{(\nu)}$ from configuration $e$ to $e'$ interacting with bath 
$\nu=L,R$ can be written as 
\begin{equation}
U_{e' e}^{(\nu)}=\frac{\displaystyle \beta_{\nu}}{\displaystyle 2}  \, \textrm{e}^{-\beta_{\nu} e'} \, , \nonumber
\end{equation}
and $U_{e' e}=U_{e' e}^{(L)} + U_{e' e}^{(R)}$. The transition rate is normalized, $\sum_{e'} U_{e' e}=1$, thus guaranteeing the conservation of probability during the stochastic 
evolution. This model can be regarded as the single-site version of the one-dimensional Kipnis-Marchioro-Presutti (KMP) model of heat conduction \cite{kmp,HG}. Having a single site 
then implies that no energy diffusion takes place, but only energy exchange with the thermal reservoirs.
Associated to the interaction with the baths there is an energy current through the system, $q_{e' e}^{(\nu)}$. We may define this current at the microscopic level in different ways. 
For the time being, let us define $q_{e' e}^{(\nu)}$ in a symmetric way
\begin{eqnarray}
\label{current1}
q_{e' e}^{(\nu)} \!=\!
\left\{ \! \begin{array}{cc}
{\displaystyle \frac{e'-e}{2} } &{\displaystyle \quad \nu=L} \phantom{\, .} \\ \\
{\displaystyle \frac{e-e'}{2} } &{\displaystyle \quad \nu=R} \, .
\end{array}
\right.
\end{eqnarray}
In this way, positive currents correspond to the transport of energy from the left to the right reservoir. In the steady state, the probability density for having an energy $e$
can be trivially calculated 
\begin{equation}
\text{P}(e)=\frac{1}{2}\left(\beta_L \text{e}^{-\beta_L e} +\beta_R \text{e}^{-\beta_R e} \right) \, , \nonumber
\end{equation}
and hence local equilibrium does not hold for this toy model. Notice that $\text{P}(e)$, despite being non-Gibbsian, can be related to the transition rate $U_{e' e}$ via detailed balance.
The mean energy is $\langle e \rangle = \frac{1}{2}(T_L+T_R)$, while the average current is $\langle q \rangle = \frac{1}{4}(T_L-T_R)$. 
Therefore Fourier's law trivially holds in this system for arbitrarily large temperature gradients, $\nabla T =\frac{1}{2}(T_L-T_R)$,
with a thermal conductivity $\kappa(T)=\frac{1}{2}$. Moreover, in equilibrium
($T_L=T_R\equiv T$), current fluctuations have a variance $\sigma(T)=T^2$. These values of $\kappa(T)$ and $\sigma(T)$ agree with the thermal conductivity and current variance 
of the spatially-extended KMP model \cite{kmp}, reinforcing the view that this model is just the single site version of the KMP process.

\subsection{Current Large Deviation Function}

Let now $\text{P}(e,q,t;e_0)$ be the probability of having the system in configuration $e$ with a total \emph{time-integrated} current $q t$ after a long time $t$,
starting from a configuration $e_0$. This probability depends also on the temperatures of the boundary thermal baths, but we drop this dependence here for notation convenience.
 $\text{P}(e,q,t;e_0)$ obeys the following master equation\footnote{All throughout the paper we 
are using the symbol $\sum_e$ in a general context, meaning ``sum over configurations'', independently of whether configuration space is discrete or continuous. In fact, 
for the model studied here configuration space is continuous, and sums become integrals over the system energy.}
\begin{equation}
\text{P}(e,q,t;e_0) = \sum_{e'} \sum_{\nu} U_{e e'}^{(\nu)} \text{P}(e',q t -q_{e e'}^{(\nu)},t-1;e_0) \, . \nonumber
\end{equation}
Iterating in time, we can write the above probability as
\begin{equation}
\text{P}(e,q,t;e_0) = \sum_{e_{t-1}\ldots e_1} \sum_{\nu_{t} \ldots \nu_1} U_{e e_{t-1}}^{(\nu_t)} \ldots U_{e_1 e_0}^{(\nu_1)}
\, \delta [q t - (q_{e e_{t-1}}^{(\nu_t)}+ \ldots + q_{e_1 e_0}^{(\nu_1)})] \, . \nonumber
\end{equation}
This is nothing but the weighted sum over all possible phase-space paths $\{e \ldots e_0 \}$ starting at $e_0$ and ending at $e$, with duration $t$, such that the total 
time-integrated current is $q t$. One may then write the probability of observing a total current $q t$ as $\text{P}(q,t;e_0)=\sum_{e} \text{P}(e,q,t;e_0)$.
For long times 
$\text{P}(q,t;e_0)$ obeys a large deviation principle \cite{LD,Tou}
\begin{equation}
\text{P}(q,t;e_0)\sim \text{e}^{+t {\cal F}(q)} \, , 
\label{ldp}
\end{equation}
where ${\cal F}(q)$ is the current large-deviation function (LDF), which does
not depend on the initial state $e_0$. 
The function ${\cal F}(q)$ is typically everywhere negative except for $q=\langle q \rangle$, for which it is zero, meaning that current fluctuations away from the average 
are exponentially unlikely in time. In most cases it is convenient to work with the moment-generating functions of the above distributions
\begin{equation}
{\Pi(e,\lambda,t;e_0) = \sum_{q} \text{e}^{t \lambda q} \text{P}(e,q,t;e_0) = \sum_{e_{t-1}\ldots e_1} \sum_{\nu_{t} \ldots \nu_1} U_{e e_{t-1}}^{(\nu_t)} 
\ldots U_{e_1 e_0}^{(\nu_1)} \, \text{e}^{\lambda (q_{e e_{t-1}}^{(\nu_t)}+ \ldots + q_{e_1 e_0}^{(\nu_1)} )}} \, ,
\label{pi1}
\end{equation}
and $\Pi(\lambda,t;e_0) =\sum_{e} \Pi(e,\lambda,t;e_0)$.
For long $t$, one can show that $\Pi(\lambda,t;e_0) \sim \text{e}^{t \mu(\lambda)}$, where $\mu(\lambda)= \max_q [{\cal F}(q) + \lambda q]$ is the Legendre transform of the current LDF. 
We can now define a modified dynamics, $\tilde{U}_{e' e}^{(\nu)} \equiv \text{e}^{\lambda q_{e' e}^{(\nu)}}\, U_{e' e}^{(\nu)}$, so
\begin{equation}
\Pi(\lambda,t;e_0)  = \sum_{e \ldots e_1} \sum_{\nu_{t} \ldots \nu_1} \tilde{U}_{e e_{t-1}}^{(\nu_t)} \ldots \tilde{U}_{e_1 e_0}^{(\nu_1)}  \, .
\label{pi2}
\end{equation}
This dynamics is however not normalized. For our particular model
\begin{eqnarray}
\label{modrateSS}
\tilde{U}_{e' e}^{(\nu)} \!=\!
\left\{ \! \begin{array}{cc}
{\displaystyle  \frac{\beta_L}{2} \, \text{e}^{-(\beta_L-\frac{\lambda}{2}) e'} \, \text{e}^{-\frac{\lambda}{2} e} } &{\displaystyle \quad \nu=L } \phantom{\, ,} \\ \\
{\displaystyle  \frac{\beta_R}{2} \, \text{e}^{-(\beta_R+\frac{\lambda}{2}) e'} \, \text{e}^{\frac{\lambda}{2} e} } &{\displaystyle \quad \nu=R } \, ,
\end{array}
\right.
\end{eqnarray}
The simplicity of this single-site model allows the direct calculation of the \emph{sum} (\ref{pi2}). Let us start by defining the following recurrence
\begin{equation}
X_t^{(\nu)}(e;e_0)=\sum_{e_{t-1}} \sum_{\nu_{t-1}} \tilde{U}_{e e_{t-1}}^{(\nu)} X_{t-1}^{(\nu_{t-1})}(e_{t-1};e_0) \, , \nonumber
\end{equation}
with $X_1^{(\nu)}(e;e_0)\equiv \tilde{U}_{e e_0}^{(\nu)}$. In this way, eq. (\ref{pi2}) can be written as
\begin{equation}
\Pi(\lambda,t;e_0)=\sum_{e} \sum_{\nu} X_t^{(\nu)}(e;e_0) \, .
\label{pi3}
\end{equation}
A simple induction argument shows that
\begin{equation}
X_t^{(R)}(e;e_0)= A_t^{(R)}(e_0) \, \text{e}^{-(\beta_R+\frac{\lambda}{2}) e} \quad ; \quad X_t^{(L)}(e;e_0)= A_t^{(L)}(e_0) \, \text{e}^{-(\beta_L-\frac{\lambda}{2}) e} \, , \nonumber
\end{equation}
where  $A_t^{(R,L)}(e_0)$ are some coefficients, and
$-\beta_R<\lambda<\beta_L$ for the integrals to converge. 
Plugging this into into eq. (\ref{pi3}) we arrive at
\begin{equation}
\Pi(\lambda,t;e_0) = \frac{A_t^{(R)}(e_0)}{\beta_R +  \frac{\lambda}{2}} + \frac{A_t^{(L)}(e_0)}{\beta_L -  \frac{\lambda}{2}} \, .
\label{pi4}
\end{equation}
The coefficients $A_t^{(\nu)}(e_0)$ obey the following recurrence (in matrix form)
\begin{equation}
\vec{A}_{t+1} \equiv \left( \begin{array}{c}{\displaystyle A_{t+1}^{(R)}} \\ 
{\displaystyle A_{t+1}^{(L)}}
\end{array} \right) \equiv \hat{S} \left( \begin{array}{c}{\displaystyle A_{t}^{(R)}} \\ 
{\displaystyle A_{t}^{(L)}} \end{array} \right)
= \frac{1}{2} \left( \begin{array}{cc}
{\displaystyle 1} & {\displaystyle \frac{\beta_R}{\beta_L-\lambda} } \\
{\displaystyle \frac{\beta_L}{\beta_R+\lambda} } & {\displaystyle  1 }
\end{array} \right) \, 
\left( \begin{array}{c}{\displaystyle A_{t}^{(R)}} \\ 
{\displaystyle A_{t}^{(L)}}
\end{array} \right) \, , \nonumber
\end{equation}
\begin{figure}
\centerline{
\psfig{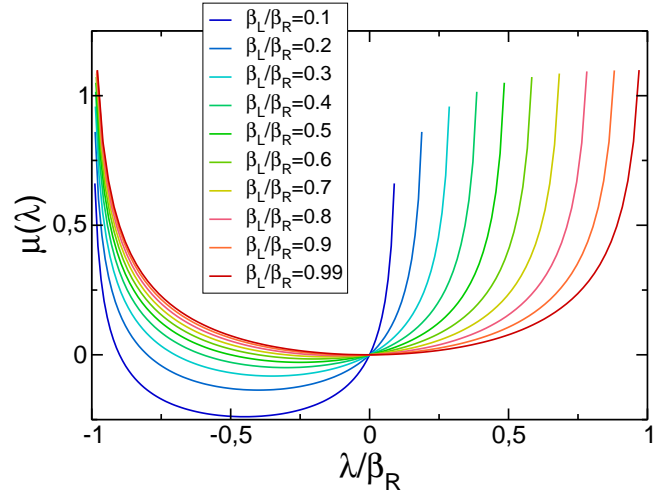}}
\caption{Legendre transform of the current LDF, $\mu(\lambda)=\max_q[{\cal F}(q)+\lambda q]$, as a function of $\lambda/\beta_R$ 
for different values of $\beta_L/\beta_R$, see eq. (\ref{muSS}). The domain of definition of $\mu(\lambda)$ is $-\beta_R < \lambda < \beta_L$. 
}
\label{muSSth}
\end{figure}
with $A_1^{(R)}(e_0)=\beta_R \, \text{e}^{\frac{\lambda}{2} e_0}/2$ and $A_1^{(L)}(e_0)=\beta_L \, \text{e}^{-\frac{\lambda}{2} e_0}/2$.
The recurrence matrix $\hat{S}$ has eigenvalues $\alpha_{\pm}=[1\pm \sqrt{\frac{\beta_R \beta_L}{(\beta_R+\lambda)(\beta_L-\lambda)}}]/2$, 
$\alpha_+>\alpha_-$, and associated eigenvectors $\vec{\psi}_{\pm}^{\text{T}} = [1,\pm \phi(\lambda)]$, with 
\begin{equation}
\phi(\lambda)\equiv \sqrt{\frac{\beta_L(\beta_L-\lambda)}{\beta_R(\beta_R+\lambda)}} \, . \nonumber
\end{equation}
Using these eigenvectors as a basis, we write $\vec{A}_1(e_0)=a_+(e_0) \vec{\psi}_+ + a_-(e_0) \vec{\psi}_-$, so 
$\vec{A}_t=\hat{S}^{t-1} \vec{A}_1=a_+ \alpha_+^{t-1 }\vec{\psi}_+ + a_- \alpha_-^{t-1 }\vec{\psi}_-$. For long times the largest eigenvalue ($\alpha_+$) dominates,
so $A_t^{(R)} \sim a_+ \alpha_+^{t-1}$ and $A_t^{(L)} \sim a_+ \phi(\lambda) \alpha_+^{t-1}$ in this limit. Using these long time asymptotics in eq. (\ref{pi4}), and recalling that 
$\Pi(\lambda,t;e_0)\sim\exp[t \mu(\lambda)]$ for long times, we find
\begin{equation}
\mu(\lambda) = \ln \left\{ \frac{1}{2}\left[ 1+ \sqrt{\frac{\beta_R \beta_L}{(\beta_R+\lambda)(\beta_L-\lambda)}} \, \right] \right\} \, .
\label{muSS}
\end{equation}

Fig. \ref{muSSth} shows $\mu(\lambda)$ versus $\lambda/\beta_R$ for different
values of the ratio $\beta_L/\beta_R$. For all pairs $(\beta_L,\beta_R)$, we
have $\mu(0)=0$ due to the normalization of $\text{P}(q,t;e_0)$, see
the definition of $\Pi(\lambda,t;e_0)$ in eq. (\ref{pi1}) and discussion
below. One can also check that the first derivative of
$\mu(\lambda)$ evaluated at $\lambda=0$ yields the average current \cite{Tou},
i.e. $\mu'(0)=\la q\ra = \frac{1}{4}(T_L-T_R)$. In addition, as a result of the time reversibility of microscopic dynamics,  
the Gallavotti-Cohen fluctuation relation \cite{GC,LS,K} holds for this system,
$\mu(\lambda)=\mu(-\lambda-E)$ with $E=\beta_R-\beta_L$. This relation
for $\mu(\lambda)$ is equivalent to the usual fluctuation relation 
${\cal  F}(-q)={\cal F}(q) - Eq$ for the current LDF \cite{Derrida,GC,LS,K}.

\newpage

\section{Statistics for a Large Deviation Event}

We now study the system energy distribution in a large deviation event of (long) duration $t$ and time-integrated current $q t$. One may measure this
distribution at a time $\tau$ such that: (i) $1 \ll \tau \ll t$, i.e. for intermediate times, or (ii) for $\tau=t$, i.e. at the end of the large deviation event. We first focus on endtime statistics.
Consider then the probability $\text{P}_{\text{end}}(e|q;t,e_0)\equiv \text{P}(e,q,t;e_0)/\text{P}(q,t;e_0)$ of having a configuration $e$ \emph{at the end} of a large deviation 
event associated to a current $q$. One can easily show in the long time limit that 
$\text{P}_{\text{end}}(e|\lambda;t,e_0) \equiv \Pi(e,\lambda,t;e_0)/\Pi(\lambda,t;e_0) = \text{P}_{\text{end}}[e|q^*(\lambda);t,e_0]$, where $q^*(\lambda)$ is the 
current conjugate to parameter $\lambda$, such that ${\cal F}'(q^*)+\lambda=0$, with $\Pi(\lambda,t;e_0)$ defined in eq. (\ref{pi4}) and 
$\Pi(e,\lambda,t;e_0)=X_t^{(R)}(e;e_0) + X_t^{(L)}(e;e_0)$. 
For long times $\text{P}_{\text{end}}(e|\lambda;t,e_0)$ converges to the following distribution, independent of both $t$ and the initial state $e_0$,
\begin{equation}
\text{P}_{\text{end}}(e|\lambda) = R \times \left[ \text{e}^{-(\beta_R+\frac{\lambda}{2})e} + \phi(\lambda) \text{e}^{-(\beta_L-\frac{\lambda}{2})e}\right] \, ,
\label{PendSS}
\end{equation}
where $R$ is a normalization constant, and we have used that $A_t^{(L)}/A_t^{(R)}\to \phi(\lambda)$ for $t \gg 1$. 
\begin{figure}
\centerline{
\psfig{file=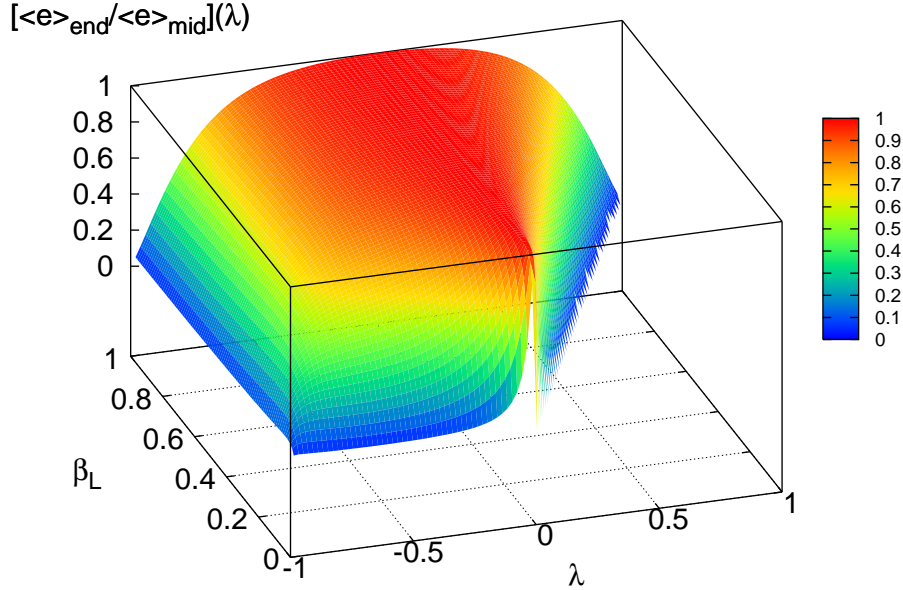,width=9cm,angle=-90}}
\caption{Ratio of average energy at the end of a large deviation event and for intermediate times, for different values of $\lambda\in[-\beta_R,\beta_L]$ and $\beta_L\in[0,1]$, 
with fixed $\beta_R=1$. 
}
\label{eratio}
\end{figure}

To compute the energy distribution for intermediate times, $1\ll \tau\ll t$, let $\bar{P}(e,q,\tau,t;e_0)$ be 
the probability that the system was in configuration $e$ at time $\tau$ when at time $t$ the total current is $q t$, starting from configuration $e_0$.
This probability can be written as
\begin{equation}
\bar{P}(e,q,\tau,t;e_0) = \sum_{e_{t}\ldots e_{\tau+1} e_{\tau-1} \ldots  e_1} \sum_{\nu_t \ldots \nu_1}  U_{e_t e_{t-1}}^{(\nu_t)} \cdots U_{e_{\tau+1} e}^{(\nu_{\tau+1})} \,
U_{e e_{\tau-1}}^{(\nu_{\tau})} \cdots U_{e_1 e_0}^{(\nu_1)} \, \delta [q t - (q_{e_t e_{t-1}}^{(\nu_t)}+ \ldots 
+q_{e_1 e_0}^{(\nu_1)})] \, , \nonumber
\end{equation}
where we do not sum over $e$. Defining the moment-generating function of the above distribution, 
$\bar{\Pi}(e,\lambda,\tau,t;e_0)=\sum_{q} \exp (t \lambda q) \bar{P}(e,q,\tau,t;e_0)$, 
we can again check that the probability weight of configuration $e$ at intermediate
time $\tau$ during a large deviation event of current $q$, $\text{P}_{\text{mid}}(e|q;\tau,t,e_0) \equiv \bar{P}(e,q,\tau,t;e_0)/\text{P}(q,t;e_0)$, is also given by 
$\bar{\Pi}(e,\lambda,\tau,t;e_0)/\Pi(\lambda,t;e_0)$ for long times such that $1 \ll \tau \ll t$, with $q=q^*(\lambda)$. Now,
\begin{equation}
\bar{\Pi}(e,\lambda,\tau,t;e_0)= \sum_{e_{t}\ldots e_{\tau+1} e_{\tau-1} \ldots  e_1} \sum_{\nu_t \ldots \nu_1} \tilde{U}_{e_t e_{t-1}}^{(\nu_t)} \cdots 
\tilde{U}_{e_{\tau+1} e} ^{(\nu_{\tau+1})} \, \tilde{U}_{e e_{\tau-1}}^{(\nu_{\tau})} \cdots \tilde{U}_{e_1 e_0}^{(\nu_1)} \,
\label{pimid}
\end{equation}
so we can write $\text{P}_{\text{mid}}(e|\lambda;\tau,t,e_0)= \Pi(\lambda,t-\tau;e) \times \Pi(e,\lambda,\tau;e_0) / \Pi(\lambda,t;e_0)$. Therefore, 
using eq. (\ref{pi4}) in the limit of long times $t$ and $\tau$, such that $1 \ll \tau \ll t$, we arrive at
\begin{equation}
\text{P}_{\text{mid}}(e|\lambda) = R' \times \left[\text{e}^{-\beta_R e} + \phi(\lambda) \text{e}^{-(\beta_L-\lambda)e}  + \phi(-\lambda-E) \text{e}^{-(\beta_R+\lambda)e} 
+ \frac{\beta_L}{\beta_R} \text{e}^{-\beta_L e} \right] \, ,
\label{PmidSS}
\end{equation}
which again does not depend on $e_0$ and on times $t$ and $\tau$ (as far as $\tau \ll t$). Here $R'$ is another normalization constant and we have used the fact that 
$a_+(e)=(\frac{\beta_R}{4} \text{e}^{\frac{\lambda}{2} e} + \frac{\beta_L}{4\phi(\lambda)} \text{e}^{-\frac{\lambda}{2} e})$. 
Using eqs. (\ref{PendSS}) and (\ref{PmidSS}), we obtain the system average energy at the end of the large deviation event and for intermediate times,
\begin{eqnarray}
\la e \ra_{\text{end}} (\lambda) & = &  \frac{\displaystyle (\beta_L-\frac{\lambda}{2})^2 + \phi(\lambda) (\beta_R+\frac{\lambda}{2})^2}
{\displaystyle (\beta_L-\frac{\lambda}{2}) (\beta_R+\frac{\lambda}{2}) \left[ (\beta_L-\frac{\lambda}{2}) + \phi(\lambda) (\beta_R+\frac{\lambda}{2}) \right]  } \label{emedSSend} \\
\la e \ra_{\text{mid}} (\lambda) & = & \frac{1}{4 \alpha_+(\lambda)} \left[ T_R + \frac{\beta_R}{(\beta_L-\lambda)^2} \, \phi(\lambda) 
+ \frac{\beta_L}{(\beta_R+\lambda)^2} \, \frac{1}{\phi(\lambda)} + T_L \right]   \label{emedSSmid} \, .
\end{eqnarray}
Fig. \ref{eratio} shows the ratio $[\la e\ra_{\text{end}}/\la e\ra_{\text{mid}}](\lambda)$ as a function of $\lambda$ and $\beta_L$, for $\beta_R=1$. This ratio is typically
smaller than 1, except for $\lambda \approx 0$ (i.e. small fluctuations around the average current). Therefore the average system energy at the end of a large deviation event
systematically underestimates the average energy for intermediate times. This results from the slower exponential decay of $\text{P}_{\text{mid}}(e|\lambda)$ for large energies
as compared to $\text{P}_{\text{end}}(e|\lambda)$. 

We could do again all calculations using a different microscopic definition of the current $q_{e' e}$, see eq. (\ref{current1}). As far as the internal energy of the system 
remains bounded, the difference between the time-integrated currents obtained with different definitions of $q_{e' e}$ will be also bounded, thus guaranteeing the same long-time 
average current \cite{Derrida}. Therefore any observable obeying the Gallavotti-Cohen symmetry should be independent of the microscopic definition of the current. 
For instance, defining the current as the energy exchanged with the left heat bath
\begin{eqnarray}
\label{currentSS2}
q_{e' e} \!=\!
\left\{ \! \begin{array}{cc}
{\displaystyle e'-e } &{\displaystyle \quad \text{Left bath}} \\ \\
{\displaystyle 0 } &{\displaystyle \quad \text{Right bath}} \, ,
\end{array}
\right.
\end{eqnarray}
we obtain the same results for the large deviation function, eq. (\ref{muSS}), as expected. Moreover, results for the midtime energy statistics 
do not change for the new current definition. However, results at the end of the large deviation event do change with respect to the symmetric current definition, 
eq. (\ref{current1}). In particular, we obtain for the \emph{left} current (\ref{currentSS2})
\begin{eqnarray}
\text{P}_{\text{end}}(e|\lambda) & = & R \times \left[ \text{e}^{-\beta_R e} + \phi(\lambda) \text{e}^{-(\beta_L-\frac{\lambda}{2})e}\right]  \nonumber \\
\la e \ra_{\text{end}} (\lambda) & = &  \frac{\displaystyle (\beta_L-\frac{\lambda}{2})^2 + \phi(\lambda) \beta_R^2}
{\displaystyle (\beta_L-\frac{\lambda}{2}) \beta_R \left[ (\beta_L-\frac{\lambda}{2}) + \phi(\lambda) \beta_R \right]  } \, , \nonumber
\label{eqnarray}
\end{eqnarray}
which should be compared with eqs. (\ref{PendSS}) and (\ref{emedSSend}), respectively. This sensitivity of endtime statistics to the microscopic definition of the current is 
a reflection of the violation of the Gallavotti-Cohen symmetry for this distribution (see below).

\section{Time Reversibility and Relation Between $\text{P}_{\text{mid}}(e|\lambda)$ and $\text{P}_{\text{end}}(e|\lambda)$}

Direct inspection of eq. (\ref{PmidSS}) reveals that $\text{P}_{\text{mid}}(e|\lambda)=\text{P}_{\text{mid}}(e|-\lambda-E)$, with $E=\beta_R-\beta_L$, or equivalently 
$\text{P}_{\text{mid}}(e|q)=\text{P}_{\text{mid}}(e|-q)$, so system statistics at intermediate times during a large deviation event does not depend on the sign of the current. 
This implies in particular that $\la e^n \ra_{\text{mid}}(q)=\la e^n
\ra_{\text{mid}}(-q)$ $\forall n$. This counterintuitive symmetry of $\text{P}_{\text{mid}}(e|q)$ is a reflection of the time 
reversibility of microscopic dynamics, and it's a general result for systems obeying the local detailed balance condition \cite{LS}. 
For our particular system this condition reads $U_{e e'} p_{\text{eq}}(e') =
U_{e' e} p_{\text{eq}}(e) \text{e}^{-E q_{e' e}}$, where $p_{\text{eq}}(e)=\frac{\beta_R+\beta_L}{2} \exp [-\frac{\beta_R+\beta_L}{2} e]$ is 
an effective equilibrium weight for configuration $e$. Local detailed balance then implies a general symmetry between the forward 
modified dynamics for a current fluctuation and the time-reversed modified
dynamics for the negative current fluctuation, i.e. $\tilde{U}_{e e'}(\lambda)=p_{\text{eq}}^{-1}(e') \tilde{U}_{e' e}(-\lambda-E) p_{\text{eq}}(e)$, or in matrix form
\begin{equation}
\tilde{U}^{\text{T}}(\lambda)= \mathbf{P}_{\text{eq}}^{-1} \tilde{U}(-\lambda-E) \mathbf{P}_{\text{eq}} \, ,
\label{matrixsym}
\end{equation}
where $\tilde{U}^{\text{T}}$ is the transpose of $\tilde{U}$ and $\mathbf{P}_{\text{eq}}$ is a diagonal \emph{matrix} with components $p_{\text{eq}}(e)$ \cite{Rakos}. 
Introducing Dirac's notation\footnote{Dirac's bra and ket notation is useful in the context of the quantum Hamiltonian formalism for the master equation \cite{sim,Rakos,schutz,schutz2}. 
The idea is to assign to each system configuration $e$ a vector $|e\ra$ in phase space, which together with its transposed vector $\la e |$, form 
an orthogonal basis of a complex space and its dual \cite{schutz,schutz2}. For instance, if the number of configurations available to our system is finite 
(which is not the case in this paper), one could write $|e\ra^T = \la e|=(\ldots 0 \ldots 0, 1, 0 \ldots 0 \ldots )$, 
i.e. all components equal to zero except for the component corresponding to configuration $e$, which is $1$. In this notation, 
$\tilde{U}_{e' e}= \la e' | \tilde{U} | e \ra$, and a probability distribution can be written as a probability vector $| P(t) \ra = \sum_e P(e,t) |e\ra$,
where $P(e,t)=\la e| P(t) \ra$ with the scalar product $\la e' | e\ra = \delta(e'-e)$. If $\la s|=(1\ldots 1)$, normalization then implies $\la s|P(t)\ra =1$.} 
we can write the spectral decomposition of the modified dynamics \cite{sim,Rakos,schutz,schutz2}, 
$\tilde{U}(\lambda)=\sum_j \text{e}^{\Lambda_j(\lambda)} |\Lambda_j^R(\lambda) \ra \la \Lambda_j^L(\lambda) |$, where we assume that a complete biorthogonal 
basis of right and left eigenvectors for matrix $\tilde{U}(\lambda)$ exists, $\tilde{U} |\Lambda_j^R(\lambda) \ra = \text{e}^{\Lambda_j(\lambda)} |\Lambda_j^R(\lambda) \ra$ and 
$\la \Lambda_j^L(\lambda)| \tilde{U} = \text{e}^{\Lambda_j(\lambda)} \la \Lambda_j^L(\lambda)|$. Denoting $\text{e}^{\Lambda(\lambda)}$ the largest eigenvalue of $\tilde{U}(\lambda)$,
with associated right and left eigenvectors $|\Lambda^R(\lambda) \ra$ and $\la \Lambda^L(\lambda)|$, respectively, we find for long times
\begin{equation}
\Pi(e,\lambda,t;e_0) = \la e | \tilde{U}^t | e_0 \ra \xrightarrow{t \gg 1} \text{e}^{t \Lambda(\lambda)}  \langle \Lambda^L(\lambda) | e_0 \rangle 
\langle e | \Lambda^R(\lambda) \rangle \, .
\label{muasymp}
\end{equation}
In this way we have $\mu(\lambda)=\Lambda(\lambda)$, i.e. the Legendre transform of the current LDF is given by the natural logarithm of the largest 
eigenvalue of $\tilde{U}(\lambda)$. Since eq. (\ref{matrixsym}) implies that all eigenvalues of $\tilde{U}(\lambda)$ and $\tilde{U}(-\lambda-E)$ 
are equal, and in particular the largest, we find that $\mu(\lambda)=\mu(-\lambda-E)$ and this proves the Gallavotti-Cohen fluctuation relation.
Now, eq. (\ref{muasymp}) also implies that $\text{P}_{\text{end}}(e|\lambda) \propto \langle e | \Lambda^R(\lambda) \rangle$ for long times, so the
right eigenvector $| \Lambda^R(\lambda) \rangle$ associated to the largest eigenvalue of \emph{matrix} $\tilde{U}(\lambda)$ gives the probability of having a configuration $e$ at
the end of the large deviation event. In a similar way, one can easily show from eq. (\ref{pimid}) that, in the long time limit,
\begin{equation}
\text{P}_{\text{mid}}(e|\lambda) \propto \langle \Lambda^L(\lambda) | e \rangle \langle e | \Lambda^R(\lambda) \rangle \, . \nonumber 
\end{equation}
In general, the eigenvectors $|\Lambda^R(\lambda)\ra$ and $\la \Lambda^L(\lambda)|$ are different because $\tilde{U}$ is not symmetric. In order to compute the left eigenvector, notice that 
$|\Lambda^L(\lambda) \ra$ is the \emph{right} eigenvector of the transpose \emph{matrix} $\tilde{U}^{\text{T}}(\lambda)$ with eigenvalue $\text{e}^{\Lambda(\lambda)}$. 
Therefore, using the symmetry relation eq. (\ref{matrixsym}), one can show that if $|\Lambda^R(-\lambda-E)\ra$ is the right eigenvector of $\tilde{U}(-\lambda-E)$ with largest eigenvalue, 
such that $|\Lambda^R(-\lambda-E)\ra = \sum_e \la e | \Lambda^R(-\lambda-E)\ra |e\ra$, then
\begin{equation}
|\Lambda^L(\lambda)\ra = \sum_e p_{\text{eq}}^{-1}(e) \la e | \Lambda^R(-\lambda-E)\ra |e\ra \nonumber
\end{equation}
is the right eigenvector of $\tilde{U}^{\text{T}}(\lambda)$ associated to the same eigenvalue. In this way we find
\begin{equation}
\text{P}_{\text{mid}}(e|\lambda) \propto \la \Lambda^L(\lambda) | e\ra \la e | \Lambda^R(\lambda) \ra = 
p_{\text{eq}}^{-1}(e) \la e | \Lambda^R(-\lambda-E) \ra \la e | \Lambda^R(\lambda) \ra \, , \nonumber
\end{equation}
or equivalently
\begin{equation}
\text{P}_{\text{mid}}(e|\lambda) = K \times \frac{\text{P}_{\text{end}}(e|\lambda) \text{P}_{\text{end}}(e|-\lambda-E)}{p_{\text{eq}}(e)} \, ,
\label{profmid1}
\end{equation}
with $K$ some normalization constant. Hence important configurations at intermediate times are 
those with an significant probabilistic weight at the end of both the large deviation event and its time-reversed process. The above equation relates midtime and endtime
statistics, and proves that in general $\text{P}_{\text{mid}}(e|\lambda) = \text{P}_{\text{mid}}(e|-\lambda-E)$. 

Let us end this section by pointing out that the computational method of
Ref. \cite{sim}, which allows the direct evaluation of large deviation
functions in simulations (see next section), 
yields also the endtime statistics associated to a large deviation event, but this method cannot access midtime statistics (see however Ref. \cite{julien}). 
In this way, eq. (\ref{profmid1}) can be used in simulations to obtain midtime statistics from endtime histograms, as done below.

\section{Finite Size Effects during the Direct Evaluation of Large Deviation Functions}

In general, measuring in simulations the current large deviation function (or any other type of LDF) is a very difficult task. This is because, by definition, 
LDF's involve exponentially unlikely events, see eq. (\ref{ldp}). To overcome this problem, Giardin\`a, Kurchan and Peliti recently proposed an efficient computational 
scheme to measure LDF's in many particle systems \cite{sim}. This method uses the modified dynamics $\tilde{U}_{e' e}$ defined in eqs. (\ref{pi2})-(\ref{modrateSS}),
for which the rare events responsible of the large deviation of the observable are no longer rare \cite{sim,sim2}. Despite its success, this simulation method fails to
provide accurate measurements of the LDF for extreme current fluctuations \cite{HG}. In this section we show, using our toy model as an illustrative 
example, that these deviations are generally present, and can be traced back to finite-size effects associated to the direct evaluation of large deviation functions.
Furthermore, the simplicity of our toy transport model allows for an accurate estimation of the range of validity of the method of Ref. \cite{sim}, and the scaling of this 
range with the size parameter.
We also show that the Gallavotti-Cohen symmetry can be used to bound numerically this range of validity.

The direct evaluation of large deviation functions is based on the sum
\begin{equation}
\Pi(\lambda,t;e_0)  = \sum_{e_t \ldots e_1} \tilde{U}_{e_t e_{t-1}} \ldots \tilde{U}_{e_1 e_0} \, . \nonumber
\end{equation}
where $\tilde{U}_{e' e}=\tilde{U}_{e' e}^{(R)}+\tilde{U}_{e' e}^{(L)}$, see eq. (\ref{pi2}), and
$\tilde{U}_{e' e}^{(\nu)}=U_{e' e}^{(\nu)} \exp [\lambda q_{e' e}^{(\nu)}]$ is the modified dynamics.
In principle this sum over paths could be easily computed by a Monte Carlo sampling with transition rates $\tilde{U}_{e' e}$. However, 
the dynamics $\tilde{U}$ is not normalized, so we introduce the exit rates
$Y_e=\sum_{e'} \sum_{\nu} \tilde{U}_{e' e}^{(\nu)}$ and define the normalized dynamics 
$U'_{e' e}\equiv Y_e^{-1} \tilde{U}_{e' e}$. With this definition
\begin{equation}
\Pi (\lambda,t;e_0)=  \sum_{e_{t}\ldots e_1} Y_{e_{t-1}} U'_{e_t e_{t-1}} \ldots Y_{e_0} U'_{e_1 e_0} \, .
\label{pilambda}
\end{equation}
This sum over paths can be realized by considering an ensemble of $M \gg 1$ copies (or clones) of the system, evolving sequentially according 
to the following Monte Carlo scheme \cite{sim,sim2,julien}:
\begin{enumerate}
\item[I] Each copy evolves independently according to modified normalized dynamics $U'$.
\item[II] Each copy $m\in [1,M]$ (in configuration $e_t[m]$ at time $t$) is cloned with rate $Y_{e_t[m]}$. This means that, for each copy $m$, 
we generate a number $K_{e_t[m]}=\lfloor Y_{e_t[m]} \rfloor +1$ of identical clones with probability $Y_{e_t[m]} - \lfloor Y_{e_t[m]} \rfloor$, or
$K_{e_t[m]}=\lfloor Y_{e_t[m]} \rfloor$ otherwise (here $\lfloor x \rfloor$ represents the integer part of $x$). Note that if $K_{e_t[m]}=0$ the copy
may be killed and leave no offspring. This procedure gives rise to a total of $M'_t=\sum_{m=1}^M K_{e_t[m]}$ copies after cloning all of the original $M$ copies.
\item[III] Once all copies evolve and clone, the total number of copies $M'_t$ is sent back to $M$ by an uniform cloning probability $X_t=M/M'_t$.
\end{enumerate}
\begin{figure}
\centerline{
\psfig{file=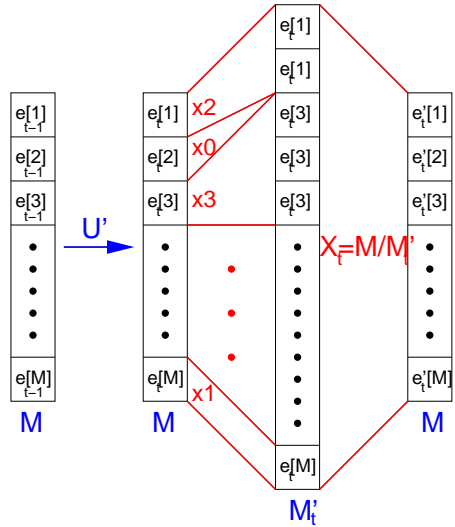,width=6cm}}
\caption{Sketch of the evolution and cloning of the copies during the direct evaluation of the large deviation function.}
\label{cloning}
\end{figure}
This process is sketched in Fig. \ref{cloning}. It then can be shown that, for long times, we recover $\mu(\lambda)$ via
\begin{equation}
\mu(\lambda)  = -\frac{1}{t} \ln \left(X_t \cdots X_0 \right)   \qquad \text{for } t\gg 1
\label{musim}
\end{equation}
To derive this expression, first consider the cloning dynamics above, but without keeping the total number of clones constant, i.e. forgetting about 
step III. In this case, for a given history $\{e_t,e_{t-1}\ldots e_1,e_0 \}$, the number ${\cal N}(e_t\ldots e_0;t)$ of copies in configuration 
$e_t$ at time $t$ obeys ${\cal N}(e_t\ldots e_0;t)= U'_{e_t e_{t-1}} Y_{e_{t-1}} {\cal N}(e_{t-1}\ldots e_0;t-1)$, so that 
\begin{equation}
{\cal N}(e_t\ldots e_0;t)=Y_{e_{t-1}} U'_{e_t e_{t-1}} \ldots Y_{e_0} U'_{e_1 e_0} {\cal N}(e_0,0) \, . \nonumber
\end{equation}
Summing over all histories of duration $t$ starting at $e_0$, see eq. (\ref{pilambda}), we find that the average total number of clones at long times shows 
exponential behavior, $\la {\cal N} (t)\ra = \sum_{e_t\ldots e_1} {\cal N}(e_t\ldots e_0;t) \sim {\cal N}(e_0,0) \exp[t \mu(\lambda)]$. Now, going back to 
step III above, when the fixed number of copies $M$ is large enough, we have $X_t = \la {\cal N} (t-1)\ra/\la {\cal N} (t)\ra$ for the global 
cloning factors, so $X_t \cdots X_1 = {\cal N} (e_0,0)/\la {\cal N} (t)\ra$ and we recover expression (\ref{musim}) for $\mu(\lambda)$.
In addition, the fraction of clones in configuration $e$ among the $M$ existing copies corresponds, for large $M$, to the ratio
$\la{\cal N}(e,t)\ra / \la{\cal N}(t)\ra$, where $\la{\cal N}(e,t)\ra
\equiv\sum_{e_{t-1}\ldots e_1} {\cal N}(e,\, e_{t-1}\ldots e_0;t)$, and this ratio is nothing
but $\text{P}_{\text{end}}(e|\lambda)$, see discussion before
eq. (\ref{PendSS}). Therefore, the fraction of clones in a given
configuration yields the system endtime statistics. Finally, using the
relation (\ref{profmid1}) between $\text{P}_{\text{mid}}(e|\lambda)$ and
$\text{P}_{\text{end}}(e|\lambda)$, we may also measure midtime statistics.
\begin{figure}
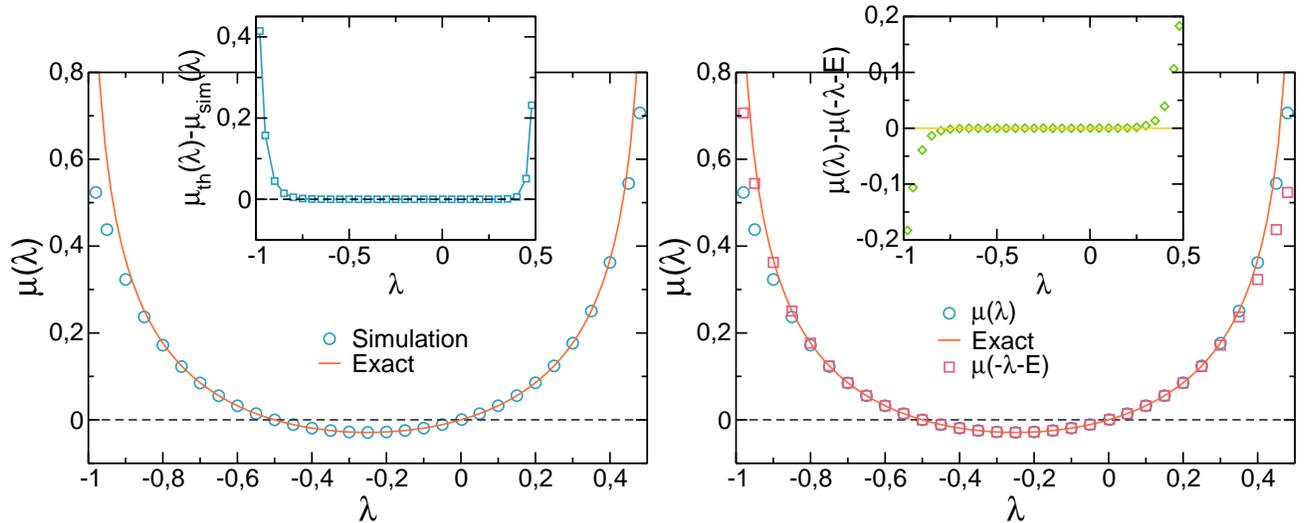

\centerline{
\psfig{file=mu-vs-lambda-single-site.eps,width=8.5cm}
\psfig{file=GC-mu-vs-lambda-single-site.eps,width=8.5cm}}
\caption{Left panel: Simulation results for $\mu(\lambda)$ and the exact
  curve. The inset shows the difference between theory and numerical results. Right panel: We test the Gallavotti-Cohen relation by plotting together
$\mu(\lambda)$ and $\mu(-\lambda-E)$, with $E=\beta_R-\beta_L$. The inset shows their difference. }
\label{muSS1}
\end{figure}

\begin{figure}
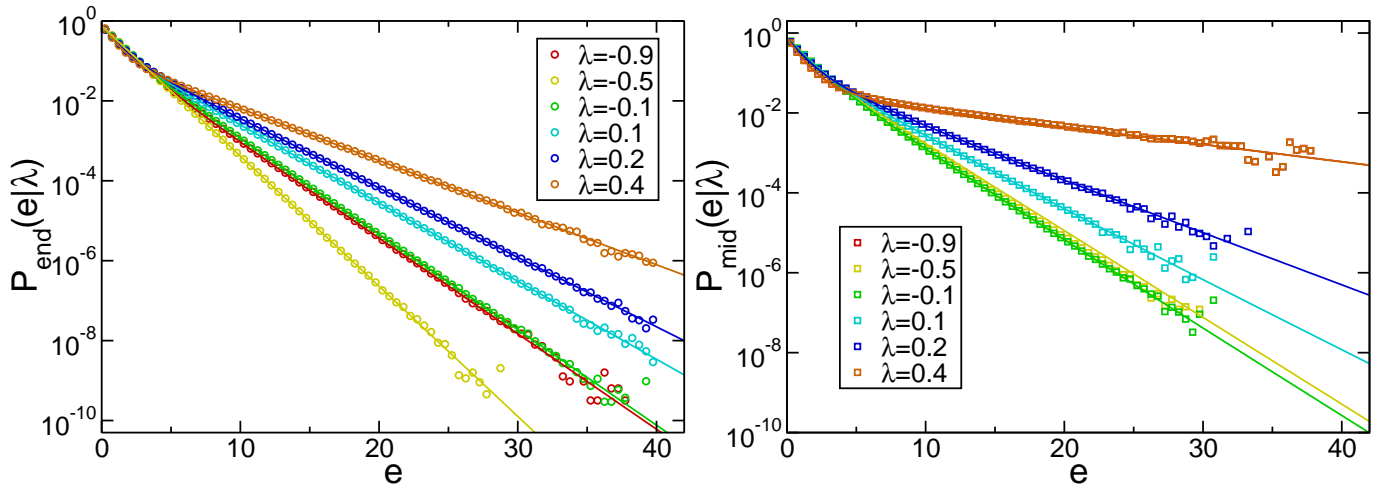

\centerline{
\psfig{file=histo-energia-varios-lambda-single-site-END.eps,width=9cm,clip}
\psfig{file=histo-energia-varios-lambda-single-site-MID.eps,width=9cm,clip}}
\vspace{0.2cm}
\caption{Left: Energy histograms measured at the end of the large deviation event for different values of $\lambda$. Points are simulation
results, and lines are exact results, eq. (\ref{PendSS}). Right: Energy histograms at intermediate times, obtained from endtime statistics via eq. (\ref{profmid1}). 
Lines are exact results from eq. (\ref{PmidSS}). }
\label{histoeSS}
\end{figure}

\begin{figure}
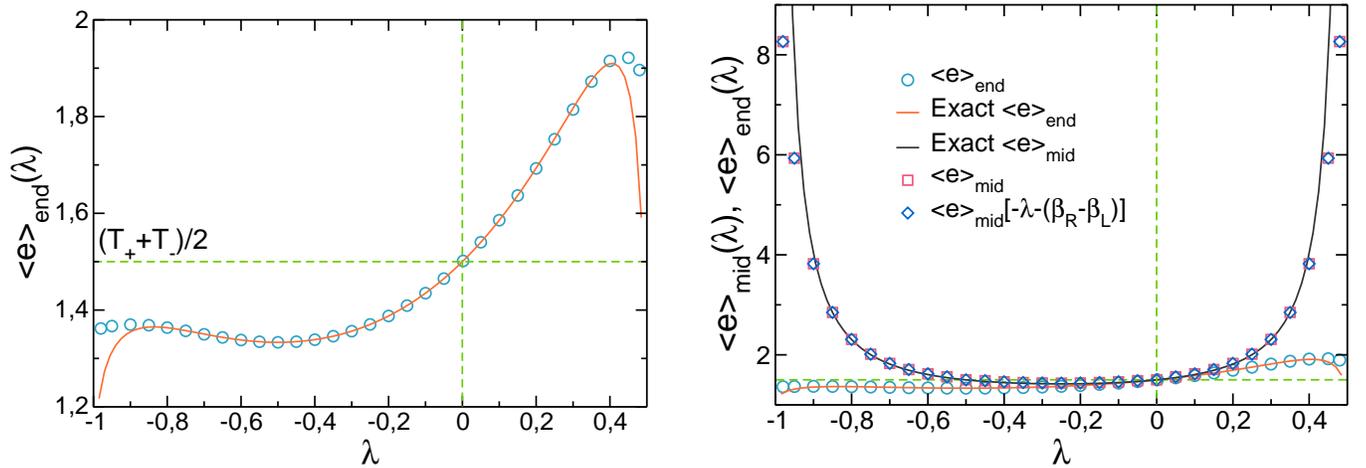

\centerline{
\psfig{file=energiaR-vs-lambda-single-site.eps,width=8.5cm} \hfill
\psfig{file=energiaI-vs-lambda-single-site.eps,width=8.5cm}}
\caption{Left: Average energy measured at the end of the large deviation event as a function of $\lambda$. Lines are exact predictions, see
eq. (\ref{emedSSend}). Right: Average energy at intermediate times, obtained from endtime statistics via eq. (\ref{profmid1}), compared with the exact result, eq. (\ref{emedSSmid}). 
$\la e \ra_{\text{mid}}(-\lambda-E)$, with $E=\beta_R-\beta_L$, is also plotted to check the symmetry of the average energy at intermediate times. 
For completeness, $\la e \ra_{\text{end}}(\lambda)$ is also shown. }
\label{RprofilesSS}
\end{figure}

We used the computational scheme described above to obtain numerical estimates
of $\mu(\lambda)$ and both $\text{P}_{\text{mid}}(e|\lambda)$ and $\text{P}_{\text{end}}(e|\lambda)$
for our toy model. In particular, we performed simulations for $T_L=2$ and
$T_R=1$, using the symmetric definition of the microscopic current, see eq. (\ref{current1}). Left panel in Fig. \ref{muSS1} shows a comparison between
simulation results for $\mu(\lambda)$ and eq. (\ref{muSS}).
Numerical measurements nicely reproduce the exact result, except for very
large current deviations, see inset to Fig. \ref{muSS1} (left).
Note that these slight differences for extreme events seem to occur earlier for
currents \emph{against the gradient}, i.e. for $\lambda<0$. On the other hand,
Fig. \ref{histoeSS} shows the energy histogram measured at the end of the large deviation event (left) and for intermediate times (right), for different 
values of $\lambda$, while Fig. \ref{RprofilesSS} shows the average endtime
(left) and midtime (right) energies. In all cases the agreement with exact results is again excellent, and 
deviations from the theoretical predictions 
are only apparent for extreme values of $\lambda$.
Our aim in what follows is to show that these deviations are due
to finite size effects in the simulation method described above \cite{sim}. 

The direct evaluation of the large deviation function is exact in the limit $M\to \infty$. For a large but finite number of copies $M$, 
this method can fail whenever the largest exit rate $Y_{e_t[m]}$ among the set of $M$ copies becomes of the order of 
$M$ itself. This condition implies that, after the cloning procedure (see Fig. \ref{cloning}), configuration $e_t[m]$ will 
overpopulate all the other copies, hence introducing a bias in the Monte Carlo sampling. We can estimate at which point this source of error becomes dominant 
by using extreme value statistics \cite{sornette}. In this way, let $Y_t^{max}\equiv \max (Y_{e_t[1]}, \ldots ,Y_{e_t[M]})$ be the maximum exit
rate among the set of $M$ copies at a given time. The probability that  $Y_t^{max}$ is smaller than a given threshold $y$ 
can be written as $\text{P}_<^{max}(y,\lambda)=[1-\text{P}_>(y,\lambda)]^M$, where $\text{P}_>(y,\lambda)$ is the probability that the exit rate of a copy is larger than $y$, 
and here we assume statistical independence among the different copies. For
large $M$, the tail of $\text{P}_>(y,\lambda)$ dominates [i.e. small values of $\text{P}_>(y,\lambda)$], and 
$\text{P}_<^{max}(y,\lambda)\simeq \exp[-M \text{P}_>(y,\lambda)]$. Therefore, the value $y^*(\lambda,p)$ of the maximum which will not be exceeded with probability $p$ 
satisfies the following equation
\begin{equation}
\text{P}_>[y^*(\lambda,p),\lambda]=\frac{1}{M} \ln(\frac{1}{p}) \, . \nonumber
\end{equation}
Now, the maximum exit rate allowed by the algorithm is $M$ itself. Setting $y^*(\lambda_c,p)=M$ in the previous identity, 
we get an equation for $\lambda_c(p)$, the critical value of $\lambda$ beyond
which the maximum which will not be exceeded with probability $p$ is larger than $M$,
\begin{equation}
\text{P}_>[M,\lambda_c(p)]=\frac{1}{M} \ln(\frac{1}{p}) \, .
\label{extremev2}
\end{equation}
This condition signals (with confidence level $p$) the onset of the systematic
bias due to the finite number of clones in simulations. 
To further proceed, we need the probability $\text{P}_>(y,\lambda)$ 
that the exit rate of a copy is larger than $y$. For that, we first recall
that the fraction of copies in a state $e$ is just $\la{\cal N}(e,t)\ra / \la{\cal N}(t)\ra=\text{P}_{\text{end}}(e|\lambda)$, 
see Section III.
Therefore the probability density for having an exit rate $Y_e=\sum_{e'} \tilde{U}_{e' e}$ during the numerical evaluation of $\mu(\lambda)$ is 
$\text{P}(Y_e,\lambda)=\sum_{e''}  \text{P}_{\text{end}}(e''|\lambda) \, \delta [Y_e - Y_{e''}]$, and 
\begin{equation}
\text{P}_>(y,\lambda)=  \int_y^{\infty} \text{d}Y_e \, \text{P}(Y_e,\lambda) =
\sum_{e''} \text{P}_{\text{end}}(e''|\lambda) \, \theta(Y_{e''}-y)\, , \nonumber
\end{equation}
where $\theta(x)$ is the Heaviside step function.
Therefore the knowledge of endtime statistics during a large deviation event
allows for an estimation of the range of validity of the simulation method of Ref. \cite{sim}.

\begin{figure}
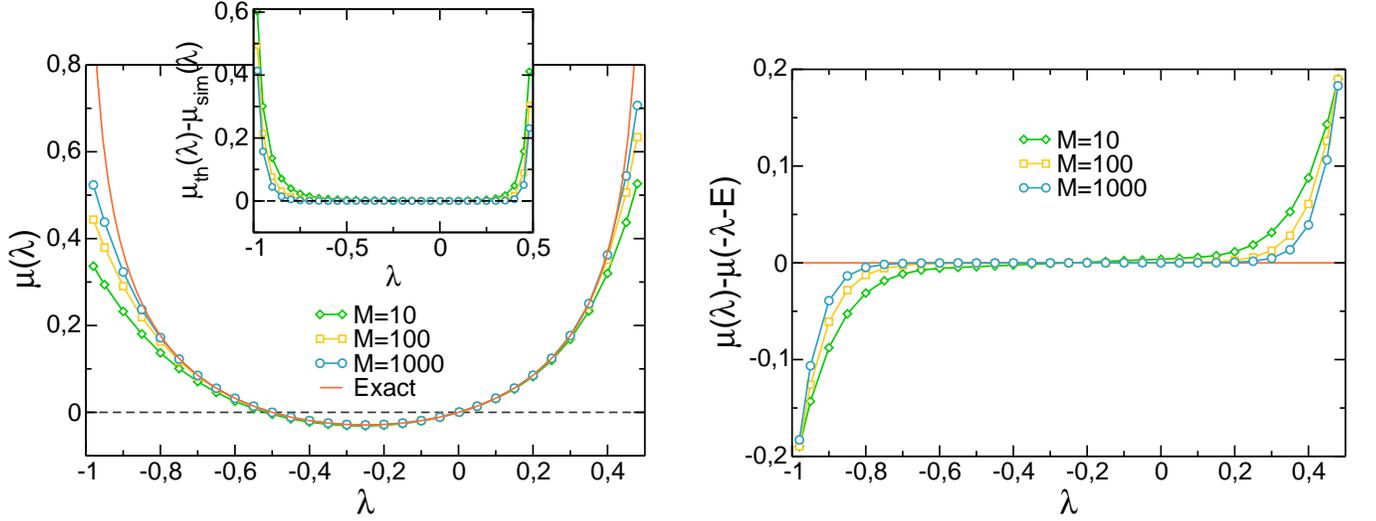

\centerline{
\psfig{file=mu-vs-lambda-single-site-FINITE-SIZE.eps,width=8.5cm} \hfill
\psfig{file=difference-mu-vs-lambda-single-site-FINITE-SIZE.eps,width=8.5cm}}
\caption{Left: Simulation results for $\mu(\lambda)$ obtained for different number of clones $M$, and exact result. 
The inset shows the difference between theory and numerical results for different $M$.
Right: Test of the Gallavotti-Cohen symmetry for different values of $M$. }
\label{finitesize}
\end{figure}

We can make these calculations explicitely in our toy transport model, for
which we know $\text{P}_{\text{end}}(e|\lambda)$, see eq. (\ref{PendSS}).
From the definition of $\tilde{U}_{e' e}$ for our model, eq. (\ref{modrateSS}), we find
\begin{equation}
Y_e = \frac{\beta_L}{2\beta_L - \lambda} \text{e}^{-\frac{\lambda}{2} e} +  \frac{\beta_R}{2\beta_R + \lambda} \text{e}^{\frac{\lambda}{2} e} \, . \nonumber
\end{equation}
Solving for $e$ in this equation
\begin{equation}
e_{\pm}(Y_e,\lambda) = \frac{2}{\lambda} \ln \left [\left(\frac{\beta_R +\lambda/2}{\beta_R}\right) \left(Y_e \pm \sqrt{Y_e^2 - 
\frac{\beta_R \beta_L}{(\beta_R+\lambda/2)(\beta_L-\lambda/2)}}\right) \right] \, , \nonumber
\end{equation}
so in principle only two possible energy configurations, $e_{\pm}$, are compatible with a given exit rate $Y_e$. 
For large values of $Y_e$, which correspond to the dominating tail of $\text{P}(Y_e,\lambda)$, there is only one solution with physical meaning (i.e. positive energy): 
$e_+(Y_e,\lambda)$ for $\lambda>0$ and $e_-(Y_e,\lambda)$ for $\lambda<0$. Therefore, by conservation of probability, 
$\text{P}(Y_e,\lambda) \, \text{d} Y_e = \text{P}_{\text{end}}(e|\lambda) \, \text{d} e$, and in the limit of large $Y_e$ we find the following asymptotic scaling
\begin{eqnarray}
\label{scalingPY}
\text{P}(Y_e,\lambda>0) \sim Y_e^{-\frac{\beta_L}{2\lambda}} \qquad ; \qquad  \text{P}(Y_e,\lambda<0) \! \sim \!
\left\{ \! \begin{array}{cc}
{\displaystyle Y_e^{-\frac{2(\beta_L-\lambda)}{|\lambda|}} } &{\displaystyle \quad -\frac{\beta_R}{2}<\lambda<0 } \\ \\
{\displaystyle Y_e^{-\frac{\beta_R}{2|\lambda|}} } &{\displaystyle \quad -\beta_R<\lambda<-\frac{\beta_R}{2} } \, .
\end{array}
\right.
\end{eqnarray}
These power laws imply also algebraic behavior for the cumulative distribution, $\text{P}_>(y,\lambda)\sim y^{-[\alpha(\lambda)-1]}$ if $\alpha(\lambda)$ 
is the exponent for $\text{P}(Y_e,\lambda)$. Therefore from eq. (\ref{extremev2}) we find $\alpha(\lambda_c)=2-\frac{\ln[\ln(p^{-1})]}{\ln(M)}$, arriving at\footnote{This calculation 
neglects the $\lambda$-dependent amplitudes of the power-laws $\text{P}(Y_e,\lambda)$ in eq. (\ref{scalingPY}). These constants give subleading corrections to the
estimation of $\lambda_c^{\pm}(p)$ derived from the $\lambda$-dependence of the power-law exponents. On the other hand, notice that in order to obtain $\lambda_c^-(p)$
we use $\alpha(\lambda)=\frac{\beta_R}{2|\lambda|}$ because the alternative exponent $\alpha(\lambda)=\frac{2(\beta_L-\lambda)}{|\lambda|}$ yields values of 
$\lambda_c^-(p)$ outside its domain, $-\frac{\beta_R}{2}<\lambda<0$.}
\begin{equation}
\lambda_c^+(p) = \frac{\beta_L}{1-\frac{\displaystyle 2\ln[\ln(p^{-1})]}{\displaystyle \ln(M)}} \xrightarrow{M \to \infty} \beta_L \qquad ; \qquad 
\lambda_c^-(p) = - \frac{\beta_R}{1-\frac{\displaystyle 2\ln[\ln(p^{-1})]}{\displaystyle \ln(M)}} \xrightarrow{M \to \infty} - \beta_R \, .
\label{lcrit}
\end{equation}
In this way, for a simulation with $M$ copies and $\lambda\in [-\lambda_c^-(p),\lambda_c^+(p)]$, the maximum exit rate among all the copies will be smaller than 
$M$ with a probability $p$, thus guaranteeing that (with confidence level $p$) no finite size effects will bias the result. Therefore the interval 
$[-\lambda_c^-(p),\lambda_c^+(p)]$ defines the range of validity of the method of Ref. \cite{sim}. For a high confidence level $p=0.99$ and $M=10^3$ clones
(as used in our simulations), we find $\lambda_c^+(p)\approx 0.38$ and $\lambda_c^-(p)\approx -0.75$, in excellent agreement with the observed deviation from
the exact result, see inset to left panel in Fig. \ref{muSS1}. This calculation can be repeated for different values of $M$, finding equally good agreement as far as $M$ is large enough, see 
Fig. \ref{finitesize} (left). 

An important consequence of eqs. (\ref{lcrit}) is that the range of validity of the algorithm, $[-\lambda_c^-(p),\lambda_c^+(p)]$, increases \emph{logarithmically}
with the number of clones used in the simulation. Therefore an appreciable increase of this range of validity in $\lambda$-space demands an exponential increase in the number of clones,
which is unfeasible in most situations of interest. We conjecture that this logarithmic increase of the range $[-\lambda_c^-(p),\lambda_c^+(p)]$ is not particular of 
our toy model, but a general feature for models with bounded values of $\lambda$. However, this shortcoming does not forbid in practice to push the algorithm far enough to observe
very large current deviations. Indeed, the current $q^*(\lambda)$ conjugate to parameter $\lambda$ increases very rapidly as $\lambda$ approaches its bounds (-$\beta_R$ and 
$\beta_L$ in this model), so a small (logarithmic) increase of the validity range $[-\lambda_c^-(p),\lambda_c^+(p)]$ for $\mu(\lambda)$ is translated into a large increase 
of the range of current values for which we can measure with confidence the current LDF, ${\cal F}(q)$.


For most many particle non-trivial systems of interest we usually don't know the analytical
form of the large deviation function to be measured in simulations. It is
hence necessary to develop a reliable method to determine the range of validity of our
numerical measurements. It is at this point where the
symmetries of the large deviation function play a prominent role. In
particular, most systems of interest obey the Gallavotti-Cohen fluctuation
theorem which results from the time reversibility of microscopic
dynamics. This relation can be stated as $\mu(\lambda)=\mu(-\lambda-E)$ for the
Legendre transform of the LDF, with $E$ some constant, see also
eq. (\ref{fth}). For our toy model, $E=\beta_R-\beta_L$. Violations of this
fluctuation symmetry are expected whenever finite-size corrections induce a
bias in the measurement of $\mu(\lambda)$, and therefore the Gallavotti-Cohen 
symmetry of the current LDF can be used to numerically bound the range of
validity of simulation results. This is confirmed in Figs. \ref{muSS1} and \ref{finitesize} (right panels), to be
compared with their corresponding left panels.

\section{Conclusions}

In this paper we studied a toy model of heat transport between two thermal baths at different temperatures. The model is simple enough so the exact current 
large deviation function can be analytically calculated, as well as the full system statistics associated to a large deviation event. In this way we find a
relation between system statistics at the end of the large deviation event and for intermediate times, which shows that important configurations at intermediate 
times are those with a significant probabilistic weight at the end of both the large deviation event and its time-reversed process. Thus midtime statistics 
turns out to be independent of the sign of the current, a reflection of the time-reversal symmetry of microscopic dynamics, while endtime statistics does 
depend on the current sign, and also on the microscopic definition of the current. The relation between midtime and endtime statistics is a general property 
for systems obeying the local detailed balance condition, which guarantees the time-reversibility of the dynamics. 
We also compared our exact results with simulation data to analyze the range of validity of a recently proposed computational 
scheme to directly evaluate large deviation functions. 
This comparison offers insights into the finite-size corrections associated to this simulation method. In particular, we were able to calculate the range of 
validity of the numerical results for our simplified model, finding that this range grows logarithmically with the number of clones involved in the evaluation 
of the current large deviation function. We conjecture that such slow growth, which impedes in practice obtaining reliable results for extreme current 
fluctuations, is a general feature of the simulation method. Finally, for more general systems for which we do not know the analytical form of the large deviation 
function, we show that violations of the Gallavotti-Cohen symmetry can be used to bound the range of validity of numerical results.

\acknowledgments

We thank B. Derrida, J.L. Lebowitz, V. Lecomte and J. Tailleur for illuminating discussions. Financial support 
from University of Granada and AFOSR Grant AF-FA-9550-04-4-22910 is also acknowledged.

\end{document}